\documentclass[aps,prl,twocolumn,showpacs,superscriptaddress,nofootinbib]{revtex4-1}  
\usepackage{graphicx,color}  
\usepackage{dcolumn}   
\usepackage{bm}        
\usepackage{amssymb} 
\usepackage{amsmath}
 
\newcommand{\be}{\begin{equation}}
\newcommand{\ee}{\end{equation}}
\newcommand{\bea}{\begin{eqnarray}}
\newcommand{\eea}{\end{eqnarray}}

\begin{document}

\def\eg{{\it e.g.}}
\newcommand{\nc}{\newcommand}
\nc{\rnc}{\renewcommand}
\rnc{\d}{\mathrm{d}}
\nc{\D}{\partial}
\nc{\K}{\kappa}
\nc{\bK}{\bar{\K}}
\nc{\bN}{\bar{N}}
\nc{\bq}{\bar{q}}
\nc{\vbq}{\vec{\bar{q}}}
\nc{\g}{\gamma}
\nc{\lrarrow}{\leftrightarrow}
\nc{\rg}{\sqrt{g}}
\rnc{\[}{\begin{equation}}
\rnc{\]}{\end{equation}}
\nc{\nn}{\nonumber}
\rnc{\(}{\left(}
\rnc{\)}{\right)}
\nc{\q}{\vec{q}}
\nc{\x}{\vec{x}}
\rnc{\a}{\hat{a}}
\nc{\ep}{\epsilon}
\nc{\tto}{\rightarrow}
\rnc{\inf}{\infty}
\rnc{\Re}{\mathrm{Re}}
\rnc{\Im}{\mathrm{Im}}
\nc{\z}{\zeta}
\nc{\mA}{\mathcal{A}}
\nc{\mB}{\mathcal{B}}
\nc{\mC}{\mathcal{C}}
\nc{\mD}{\mathcal{D}}
\nc{\mN}{\mathcal{N}}
\rnc{\H}{\mathcal{H}}
\rnc{\L}{\mathcal{L}}
\nc{\<}{\langle}
\rnc{\>}{\rangle}
\nc{\fnl}{f_{NL}}
\nc{\gnl}{g_{NL}}
\nc{\fnleq}{f_{NL}^{equil.}}
\nc{\fnlloc}{f_{NL}^{local}}
\nc{\vphi}{\varphi}
\nc{\Lie}{\pounds}
\nc{\half}{\frac{1}{2}}
\nc{\bOmega}{\bar{\Omega}}
\nc{\bLambda}{\bar{\Lambda}}
\nc{\dN}{\delta N}
\nc{\gYM}{g_{\mathrm{YM}}}
\nc{\geff}{g_{\mathrm{eff}}}
\nc{\tr}{\mathrm{tr}}
\nc{\oa}{\stackrel{\leftrightarrow}}
\nc{\IR}{{\rm IR}}
\nc{\UV}{{\rm UV}}

\title{From Planck data to Planck era: \\ Observational tests of Holographic Cosmology} 

\author{Niayesh Afshordi} \affiliation{Perimeter Institute for Theoretical Physics, 31 Caroline St. N., Waterloo, ON, N2L 2Y5, Canada} \affiliation{Department of Physics and Astronomy, University of Waterloo, Waterloo, ON, N2L 3G1, Canada}
\author{Claudio Corian\`{o}} \affiliation{STAG Research Centre, Highfield, University of Southampton, SO17 1BJ Southampton, UK} \affiliation{Mathematical Sciences, Highfield, University of Southampton, SO17 1BJ Southampton, UK} 
\affiliation{Dipartimento di Matematica e Fisica ``Ennio De Giorgi'', Universit\`{a} del Salento and INFN-Lecce, Via Arnesano, 73100 Lecce, Italy} 
\author{Luigi Delle Rose} \affiliation{STAG Research Centre, Highfield, University of Southampton, SO17 1BJ Southampton, UK}  \affiliation{School of Physics and Astronomy, Highfield, University of Southampton, SO17 1BJ Southampton, UK} \affiliation{Rutherford Appleton Laboratory, Chilton, Didcot, OX11 0QX, UK}
\author{Elizabeth Gould} \affiliation{Perimeter Institute for Theoretical Physics, 31 Caroline St. N., Waterloo, ON, N2L 2Y5, Canada} \affiliation{Department of Physics and Astronomy, University of Waterloo, Waterloo, ON, N2L 3G1, Canada}
\author{Kostas Skenderis} \affiliation{STAG Research Centre, Highfield, University of Southampton, SO17 1BJ Southampton, UK}  \affiliation{Mathematical Sciences, Highfield, University of Southampton, SO17 1BJ Southampton, UK}
 
\date{\today}

\begin{abstract}
We 
test a class of holographic models for the very early universe
against  cosmological observations and find that they are competitive to the standard $\Lambda$CDM model of cosmology. These models are based on three dimensional perturbative super-renormalizable Quantum Field Theory (QFT), and  while they predict a different power spectrum from the standard power-law used in $\Lambda$CDM, they still provide an excellent fit to data (within their regime of validity).  By comparing the Bayesian evidence for the models, we find that $\Lambda$CDM does a better job globally, while the holographic models provide a (marginally) better fit to data without very low multipoles (i.e. $l\lesssim 30$), where the dual QFT becomes non-perturbative. 
Observations can be used to exclude some QFT models, 
while we also find models  satisfying all phenomenological constraints: the data rules out the dual theory being Yang-Mills theory coupled to fermions only, but allows for Yang-Mills theory coupled to non-minimal scalars with quartic interactions. Lattice simulations of 3d QFT's can provide non-perturbative predictions for large-angle statistics of the cosmic microwave background, and potentially explain its apparent anomalies. 
\end{abstract}

\maketitle


Observations of the cosmic microwave background (CMB) offer a unique window into the very early Universe and Planck scale physics. 
The  standard model of cosmology, the so-called $\Lambda$CDM model, provides an excellent fit to observational data with 
just six parameter. 
Four of these parameters
 describe the composition and evolution of the Universe, while the other two are linked with the physics of the very early Universe.
These two parameters, the tilt $n_s$ and  and the amplitude $\Delta^2_{0}(q_*)$, parameterize the power spectrum of primordial curvature perturbations, 
\begin{equation} \label{power_sp}
\Delta_{{\cal R}}^2(q) = \Delta^2_{0}(q_*) \left(\frac{q}{q_*}\right)^{n_s-1},
\end{equation}
where $q_*$, the pivot, is an an arbitrary reference scale. 
This form of the power spectrum is  a good approximation for slow-roll inflationary models and has the ability to fit the CMB data well. Indeed, a near-power-law
scalar power spectrum may be considered as a success of the theory of cosmic inflation.

The theory of inflation is an effective theory.  It is based on gravity coupled to (appropriate) matter  perturbatively quantized around an accelarating FLRW background.  At sufficiently early times the
curvature of the FLRW  spacetime becomes large and the perturbative treatment is expected to break down -- in this regime we would need a full-fledged theory of quantum gravity.  One of the deepest insights about quantum gravity that emerged in recent times is that it is expected to be holographic \cite{'tHooft:1993gx, Susskind:1994vu,Maldacena:1997re}, meaning that there should be an equivalent description of the bulk physics using a quantum field theory with no gravity in one dimension less. One may thus seek to use holography to model the very early Universe.

Holographic dualities were originally developed for spacetimes with negative cosmological constant (the AdS/CFT duality)  \cite{Maldacena:1997re} and soon afterwards the extension to de Sitter and cosmology was considered \cite{Hull:1998vg, Witten:2001kn, Strominger:2001pn, Strominger:2001gp, Maldacena:2002vr}. 
In this context, the statement of the duality is that the partition function of the dual QFT  computes the wavefunction of the universe \cite{Maldacena:2002vr} and using this wavefunction cosmological observables may be obtained. Alternatively, \cite{McFadden:2009fg, McFadden:2010na, McFadden:2010vh, McFadden:2011kk, Bzowski:2011ab},  one may use  the Domain-wall/Cosmology correspondence \cite{Skenderis:2006jq}. The two approaches are equivalent \cite{Garriga:2014fda}.

Holography offers a new framework that can accommodate conventional inflation but also leads to qualitatively new models for the very early universe.
While conventional inflation corresponds to a strongly coupled QFT \cite{Maldacena:2011nz, Hartle:2012qb, Hartle:2012tv,Schalm:2012pi, Bzowski:2012ih, Mata:2012bx, Garriga:2013rpa, McFadden:2013ria, Ghosh:2014kba, Garriga:2014ema, Kundu:2014gxa, McFadden:2014nta, Arkani-Hamed:2015bza, Kundu:2015xta, Hertog:2015nia,Garriga:2015tea,  Garriga:2016poh}, 
the new models are associated with a weakly coupled QFT. 
These models correspond to a non-geometric bulk, and yet holography allows us to compute the predictions for the cosmological observables. We emphasise that the application of holography to cosmology is conjectural, the theoretical validity of such dualities is still open and different authors approach the topic in different ways. Here we seek to test these ideas against observations.

A class of non-geometric models were introduced in \cite{McFadden:2009fg} and their prediction have been worked out in \cite{McFadden:2009fg, McFadden:2010na, McFadden:2010vh, McFadden:2011kk, Bzowski:2011ab,Coriano:2012hd, Kawai:2014vxa}.
These models are based on three dimensional super-renormalizable QFT 
and they universally predict  a scalar power spectrum of the form,
\begin{equation} \label{power_hc}
\Delta_{{\cal R}}^2(q)= \frac{\Delta^2_0}{1+ (g q_*/q) \ln |q/\beta gq_*| +{\cal O}(g q_*/q)^2},
\end{equation}
where $g$ is related to the coupling constant of the dual QFT, while $\beta$ depends on the parameters of the dual QFT (see below).

The form of the power spectrum in (\ref{power_hc}) is distinctly different from (\ref{power_sp})\footnote{For small enough $g$, one may rewrite (\ref{power_hc}) in the form (\ref{power_sp}) with momentum dependent $n_s(q)$. However, as discussed \cite{McFadden:2009fg, Easther:2011wh}, the momentum dependence of  $n_s(q)$ is qualitatively different from that of slow-roll inflationary models \cite{Kosowsky:1995aa}.}. Since these are qualitatively different parametrizations, one may ask which of the two is preferred by the data. Note that this question is {\it a priori} independent of the underlying physical models that produced 
(\ref{power_sp}) and (\ref{power_hc}).
This question has already been addressed for WMAP7 data \cite{Komatsu:2010fb} in \cite{Easther:2011wh, Dias:2011in} and it was found that while the data mildly favour $\Lambda$CDM, it was insufficient to definitively discriminate between the two cases.  Since then, the Planck mission has released its data \cite{Ade:2015xua} and it is now time to revisit this issue.
We will present the main conclusions of the fit to Planck data here, referring to \cite{AGS} for a more detailed discussion.  

On the theoretical side, there has also been significant progress since \cite{Easther:2011wh}. While the form of (\ref{power_hc}) is universally fixed, the precise relation between $g$ and $\beta$ and the parameters of the dual QFT 
requires a 2-loop computation, which has now been carried out in \cite{CDS}. We can thus not only check whether (\ref{power_hc}) is compatible with CMB data, but also use the data to 
do a model selection. 

{\em Theory.}--- Following \cite{McFadden:2009fg}, we consider the dual QFT to be $SU(N)$ gauge theory coupled to scalars $\Phi^M$ and fermions $\psi^L$, where $M, L$ are flavor indices. The action is given by 
\bea 
S &=& \frac{1}{\gYM^2} \int d^3 x \, \tr \left[ \frac{1}{2} F_{ij} F^{ij} +  (\mD \Phi)^2 + 2 \bar{\psi} \mD\!\!\!\!/ \ \! \psi \right. \nn \\
&& \left.\qquad \qquad \qquad  
+ 2 \sqrt{2} \mu \cdot (\Phi \bar{\psi}  \psi) + \frac{1}{6} \lambda \cdot \Phi^4 \right] , \label{action}
\eea
where all fields, $\varphi = \varphi^a T^a$,  are in the adjoint of $SU(N)$ and $\tr T^a T^b = \frac{1}{2}  \delta^{ab}$.
$F_{ij}$ is the Yang-Mills field strength, and $\mD$ is a gauge covariant derivative. 
We use the shorthand notation $(\mD \Phi)^2 = \delta_{M_1M_2} \mD_i \Phi^{M_1} \mD^i \Phi^{M_2}$,  $\bar{\psi} \mD\!\!\!\!/  \ \! \psi = \delta_{L_1 L_2} \bar{\psi}^{L_1} \gamma^i \mD_i \psi^{L_2}$, 
$\mu \cdot (\Phi \bar{\psi}  \psi) \equiv  \mu_{M L_1 L_2} \Phi^M \bar{\psi}^{L_1} \psi^{L_2}$ and
$\lambda \cdot \Phi^4\equiv  \lambda_{M_1 M_2 M_3 M_4} \Phi^{M_1} \Phi^{M_2} \Phi^{M_3} \Phi^{M_4}$.


The holographic dictionary relates the scalar and tensor power spectra to the 2-point function of the energy-momentum tensor $T_{ij}$. For  the scalar power spectrum,
\be \label{power_sp2}
\Delta^2_{{\cal R}}(q) = \frac{1}{4 \pi^2 N^2 f(\geff^2)},
\ee
where $\geff^2(q) \equiv \gYM^2 N/q$ is the effective dimensionless 't Hooft coupling constant, $q$ is the magnitude of the momentum $\vec{q}$ and $f(\geff^2)$ is extracted from the momentum space 2-point of function 
of the trace of the energy momentum tensor, 
$\< T^i_i(\vec{p}) T^j_j(\vec{q}) \> = (2 \pi)^3 \delta(\vec{p} + \vec{q}) q^3 N^2 f(\geff^2)$. In perturbation theory, 
\be \label{2loop}
f(\geff^2) = f_0 \left[ 1 - f_1 \, \geff^2 \ln \geff^2 + f_2 \, \geff^2 + {\cal O}(\geff^4)  \right].
\ee
The function $f_0$ is determined by a 1-loop computation, while $f_1$ and $f_2$ come from 2-loops. The presence of the logarithm is due to UV and IR divergences in the computation of the 2-point function of the energy momentum tensor.   A detailed derivation of (\ref{power_sp2}) may be found in \cite{McFadden:2010na, Easther:2011wh}.
Following \cite{Easther:2011wh}, (\ref{power_hc}) and (\ref{power_sp2}-\ref{2loop}) match if: 
\be \label{match}
  g q_* = f_1 \gYM^2 N, \ln \frac{1}{\beta} = \frac{f_2}{f_1} + \ln |f_1|, \Delta_0^2 = \frac{1}{4 \pi^2 N^2 f_0}.
\ee
So,  a universal prediction of these class of theories is  the power spectrum (\ref{power_hc}), independent of the details of the 2-loop computation\footnote{This assumes $f_1 \neq 0$. A separate analysis is required, where $f_1=0$, e.g., for (\ref{action}) without gauge fields and fermions. }. 
 
The 1-loop computation was done in \cite{McFadden:2009fg, McFadden:2010na} and we here report the result of the 2-loop computation \cite{CDS} -- a summary of the computation is provided in the appendix..  
The final result is
\bea
f_0&=& \frac{1}{64} \mN_{(B)} , \quad \mN_{(B)} = 1 + \sum_{M}
(1-8 \xi_{M})^2 \\
f_1 &=& - \frac{4}{3 \pi^2} \frac{1}{\mN_{(B)}} \left(\mN_\psi -2  +2 \mN_\Phi + \frac{1}{2} \mu^2 - 48 \, \Sigma_\Phi \right) , \\
\ln \beta&=&\ln \frac{1}{|g|} -\frac{a_0}{f_1} - \frac{64/\pi^2}{ f_1 \mN_{(B)} }\Sigma_\Phi \ln \frac{N f_1}{g} \label{final_log}
\eea
where $\mN_{\Phi}$ and $\mN_\psi$ are the total number of scalars and fermions,  and
\bea
&&a_0 = -\frac{1}{24 \pi^2 \mN_{(B)}} \Big[16 +3 \pi ^2 - 56 \mN_\psi  - 4 \sum_M \mu_{MM}^2  +\nonumber \\
&& \sum_M
3 (8 \xi_M -1) (8 \left(\pi^2-16\right) \xi_M -3 \pi ^2+112 + 2 \mu^2_{M M} ) 
 \nonumber \\
&&   + \pi^2
\sum_{M_1, M_2}
\lambda_{M_1 M_1 M_2 M_2} \left(8 \xi_{M_1} - 1 \right) \left(8 \xi_{M_2} - 1 \right) \Big], \nonumber \\
&&\Sigma_\Phi = \sum_M
\xi_M^2 \left(2 +\frac{1}{2}  \mu^2_{MM} \right), \nonumber  
\eea
where $\mu_{M_1 M_2}^2 =  \sum_{L_1, L_2} \mu_{M_1 L_1 L_2} \mu_{M_2 L_2 L_1}$, $\xi_M$ is the non-minimality parameter\footnote{Non-minimal scalars on a curved background have the coupling $1/(2 \gYM^2) \sum_M \int \xi_M R (\Phi^M)^2$, where $R$ is the curvature scalar,  and this term induces an ``improvement term'' to their energy momentum tensor,
$T_{ij} = (2/\sqrt{g}) (\delta S/\delta g^{ij}) |_{g_{ij}=\delta_{ij}}$, see \cite{supplement}.} and summations over $M(L)$ are over scalars (fermions).

{\em Fitting to data.}--- 
We would like now to assess how well a power spectrum of the form (\ref{power_hc}) fits the cosmological data and compare with that of the conventional power-law power spectrum. 
Recall that $\Lambda$CDM is parametrized by six parameters, $(\Omega_{b}h^{2}, \Omega_{c}h^{2}, \theta, \tau, \Delta_0^2, n_s)$, where
$\Omega_{b}h^{2}$ and $\Omega_{c}h^{2}$ are the  baryon and dark matter densities, $\theta$ is the angular size of the sound horizon at recombination, $\tau$ is the the optical depth due to re-ionization
and $\Delta_0^2$, $n_s$ are the parameters entering in (\ref{power_sp}). 
To formalize the comparison, we define (following \cite{Easther:2011wh}) holographic cosmology (HC) as the model parametrized by  $(\Omega_{b}h^{2}, \Omega_{c}h^{2}, \theta, \tau, \Delta_0^2, g, \ln \beta)$\footnote{In \cite{Easther:2011wh} the parameter $\beta$ was incorrectly assumed to be equal one. We refitted the WMAP data and found that the global minimum is at $\beta=3.777$.}.
This model has 7 parameters so  in order to compare models with the same number of parameters  we also consider $\Lambda$CDM with running $\alpha_s = d n_s/d \ln q$. 
Note that our aim here is to compare {\it empirical models}, not the underlying physical models that lead to them. If the data selects one of the two empirical models, then this would falsify all physical models that underlie the other model.  
\begin{figure}[h]
\begin{center}
\includegraphics[width=0.48\textwidth]{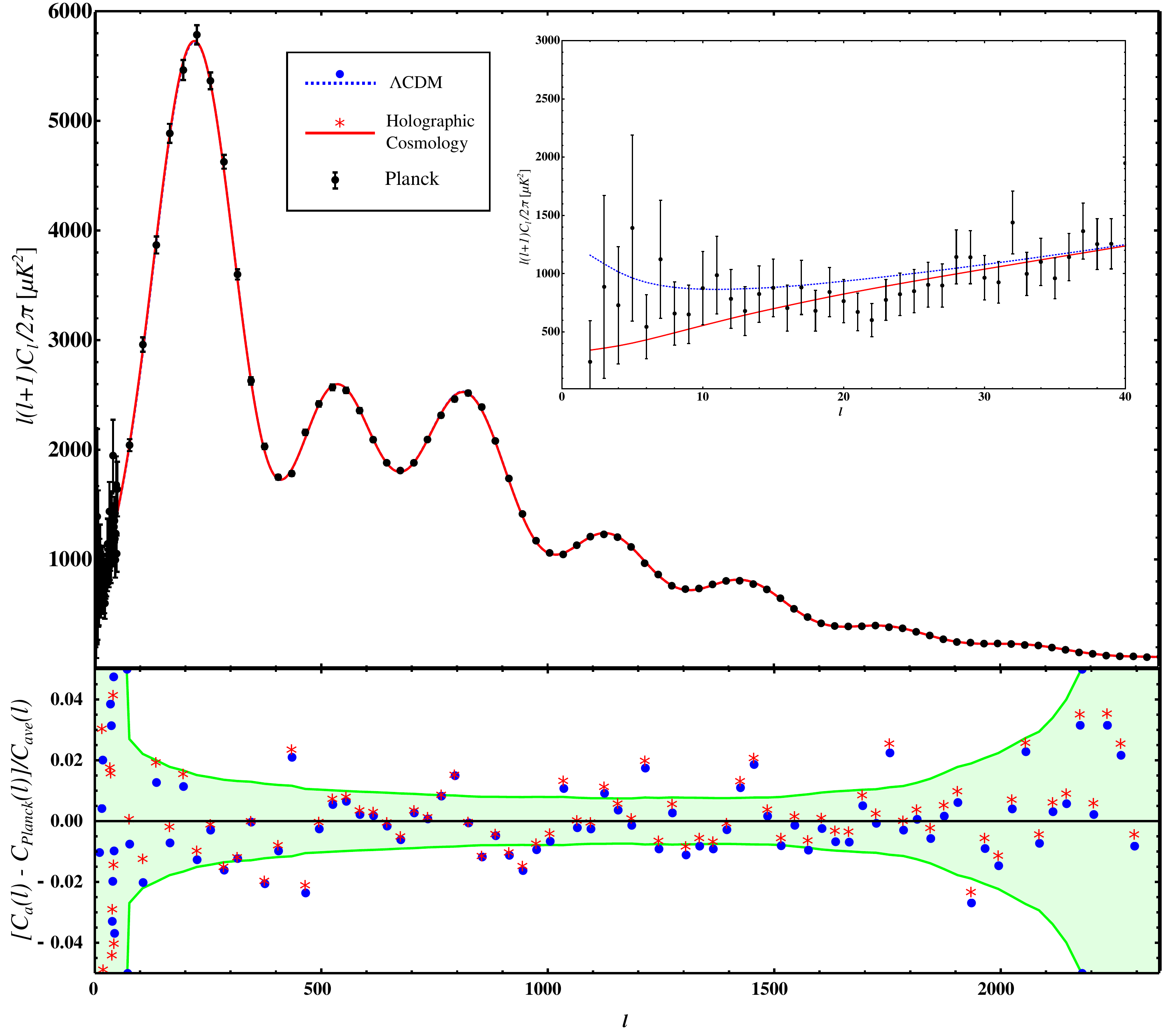}
\caption{Angular power spectrum of CMB temperature anisotropies, comparing Planck 2015 data with best fit $\Lambda$CDM (dotted/blue) and holographic cosmology (solid/red) models, for $l \geq 30$. Lower panel shows the relative residuals, where the green shaded region indicates the $68\%$ region of Planck 2015 data.
 \label{ang_sp}}
\end{center}
\end{figure} 

We analysed the data using CosmoMC \cite{Seljak:1996is,Zaldarriaga:1997va,Lewis:1999bs,Lewis:2002ah,Howlett:2012mh,Lewis:2013hha,camb_notes}.
We ran both $\Lambda$CDM and HC with the same datasets, fitting the models to the 
Planck 2015 data including lensing \cite{Ade:2015xua,Ade:2015zua,Aghanim:2015wva,Ade:2015fva,Bennett:2012zja,Reichardt:2011yv,Das:2013zf}, 
as well as Baryonic Acoustic Oscillations (BAO) \cite{Beutler:2011hx,Blake:2011en,Anderson:2012sa,Beutler:2012px,Padmanabhan:2012hf,Anderson:2013zyy,Samushia:2013yga,Ross:2014qpa} 
and BICEP2-Keck-Planck (BKP) polarization \cite{Ade:2015tva}. 
After CosmoMC had run to determine the mean and errors in the parameters, we ran the minimizer \cite{minimizer} within the code to 
determine the best fit parameters and likelihood.


The Planck angular TT spectrum  together with the best fit curves and residuals for HC and $\Lambda$CDM are presented in Fig. \ref{ang_sp}. 
Notice that the difference between $\Lambda$CDM and HC lies within the 68\% region of Planck, with the largest difference being at small multipoles. Very similar results hold for the TE and EE spectra \cite{AGS}.
We determined the best fit values  for all parameters for HC, $\Lambda$CDM and $\Lambda$CDM with running. Our values for the parameters of $\Lambda$CDM and $\Lambda$CDM with running are in agreement with those determined by the Planck team.
All common parameters of the three models are within $1\sigma$ of each other (with the notable exception of the optical depth $\tau$
\cite{AGS}). 
We report the values of $\Delta_0, g, \ln \beta$ and $\chi^2$ in Table \ref{tab:Planck-2015-3} (the list of all parameters can be found in  \cite{AGS}). 
The $\chi^2$ of the fit indicates that HC is disfavoured at about 2.2$\sigma$ relative to $\Lambda$CDM with running, when we consider all multipoles.

\begin{table}
\caption{Upper part: Planck \label{tab:Planck-2015-3}2015+BAO+BKP mean parameters for
holographic cosmology. Lower part: $\chi^2$ values for fit with all multipoles and fit with $l<30$ multipoles excluded. }
\noindent \centering{}%
\begin{tabular}{|c||c|c|c|}
\hline
HC & $10^9\Delta_0^2$   & $g$ & $\ln \beta$\tabularnewline 
\hline 
\hline 
all $l$   & $2.126^{+0.058}_{-0.058} $      & $-0.00703_{-0.00167}^{+0.00105}$    & $0.877^{+0.186}_{-0.239}$ \tabularnewline
\hline 
$l \geq 30$  & $2.044^{+0.072}_{-0.075} $     & $-0.01305_{-0.00345}^{+0.00452}$ &  $1.014^{+0.206}_{-0.272}$ \tabularnewline
\hline 
\hline
& HC & $\Lambda$CDM & $\Lambda$CDM running \tabularnewline 
\hline
\hline
$\chi^{2}$ (all $l$)       & $11324.5$ & $11319.9$ & $11319.6$ \tabularnewline
\hline 
$\chi^{2} (l \geq 30)$ & $824.0$ & $824.5$ & $823.5$ \tabularnewline
\hline 
\end{tabular}
\end{table}

\begin{figure}[h]
\begin{center}
\includegraphics[width=0.48\textwidth]{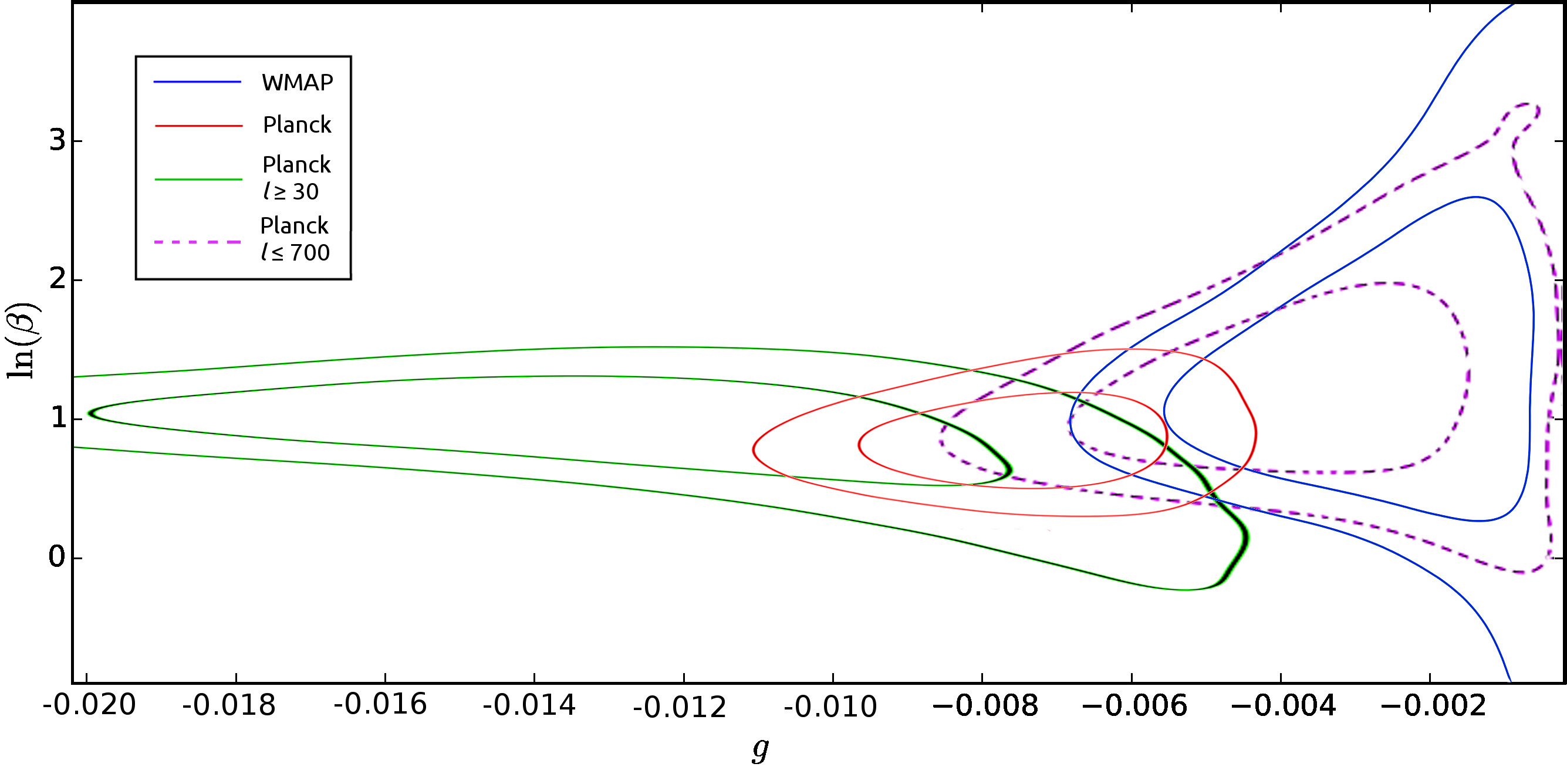}
\caption{\label{fig:g-vs-ln}Plot of $1\sigma$ and $2\sigma$ regions in parameter
space for holographic cosmology $g$ and $\ln(\beta)$ values for
WMAP (blue, right), Planck (red, middle), Planck with $l < 30$ values
removed (green, left), and Planck with $l > 700$ values ignored (purple,
dashed). We see that higher resolution data progressively pushes $g$ to lower negative values.}
\end{center}
\end{figure}

Relative to the WMAP fit in \cite{Easther:2011wh} the value of $g$ has decreased from $-1.3\times 10^{-3}$  to $-7 \times 10^{-3}$. In Fig. \ref{fig:g-vs-ln}, we investigate how the value of 
$g$ changes if we change the range of multipoles that we consider. It is clear from the plot that the value of $g$ is compatible between WMAP and Planck, if we keep the same multipoles.
It is also clear that the high $l$ modes want to push $g$ to lower negative values. Larger values of $|g|$ indicate that the theory may become non-perturbative at very low $l$  
and, as such, the predictions of the model cannot be trusted in that regime. We shall see below that this is supported by model selection criteria. Therefore, we repeat the fitting, excluding 
the $l<30$ multipoles. The results for $\Delta_0, g, \ln \beta$ and $\chi^2$ 
are tabulated in Table \ref{tab:Planck-2015-3}. With this data, all common parameters are now compatible with each other \cite{AGS}. 
The $\chi^2$ test shows that the three models are now within $1\sigma$.

The power spectrum for the tensors takes the same form as (\ref{power_hc}) but with different values of $g$ and $\beta$. We fitted the data with this form of the power spectrum and found that it is consistent with $r=0$;  the $2\sigma$ upper limit on the tensor-to-scalar ratio is $r < 0.125$.

{\em Bayesian Evidence.}--- In comparing different models, one often uses information criteria such as the value of $\chi^2$, which quantifies the goodness of a fit. 
We emphasise that with ``model'' we mean  the three empirical models introduced above, $\Lambda$CDM, $\Lambda$CDM with running and HC.
What we really want to know, however, is what is the probability for each of these models given the data. This is obtained by computing the Bayesian Evidence. 

As discussed in \cite{Easther:2011wh}, if we assume flat priors for all parameters $\alpha_M$ that define a given model, the Bayesian evidence is given by
 $E=\frac{1}{{\rm Vol}_{M}}\int d\alpha_{M}\mathcal{L\left(\alpha_{M}\right)}$,
where $\mathcal{L\left(\alpha_{M}\right)}$ is the likelihood and Vol${}_M$ is the volume of the region in parameter space over which the prior probability distribution is non-zero. 
The evidence may be computed either by using CosmoMC or by MultiNest \cite{Feroz:2007kg,Feroz:2008xx,Feroz:2013hea}.

Note that the aim here is to compare empirical models and we determined the priors from previous fits of the same empirical models to data (as is common)\footnote{Had we focused on specific physical models we could use the wavefunction of the universe to obtain corresponding theoretical priors, see \cite{Hartle:2012tv} for work in this direction.}. 
We use the priors in Table 4 of \cite{Easther:2011wh},  
except that the upper limit of 
$100\, \theta$ is taken to be 1.05. The prior for the running is taken to be $|\alpha_s| \leq 0.05$. 
The priors for $n_s$ are the asymmetric prior used in \cite{Easther:2011wh}: $0.92 \le n_s \le 1$.
For the prior for $g$ we use variable range, $g_{\rm min} \leq g < 0$. This prior is fixed by the requirement that perturbation theory is valid. We will allow for the possibility that the perturbative expansion is valid only for $l>30$. We use as a rough estimate for the validity of perturbation theory that $g q^*/q$ is sufficiently small, taking this to mean a value between 0.20 and 1 at $l=30$\footnote{The momenta and multipoles are related via $q= l/r_h$, where  $r_h=14.2$ Gpc is the comoving radius of the last scattering surface.}. This translates into $-0.009 < g_{\rm min}< -0.45$. The prior for $\beta$ is fixed by using the results from (our fit to) WMAP data. We use two sets of priors: one coming from the 1$\sigma$ range ($0 \leq \ln \beta \leq 2$)  and the other from the 2$\sigma$ range ($-0.2 \leq \ln \beta \leq 3.5$). 

\begin{figure}[h]
\begin{center}
\includegraphics[width=0.45\textwidth]{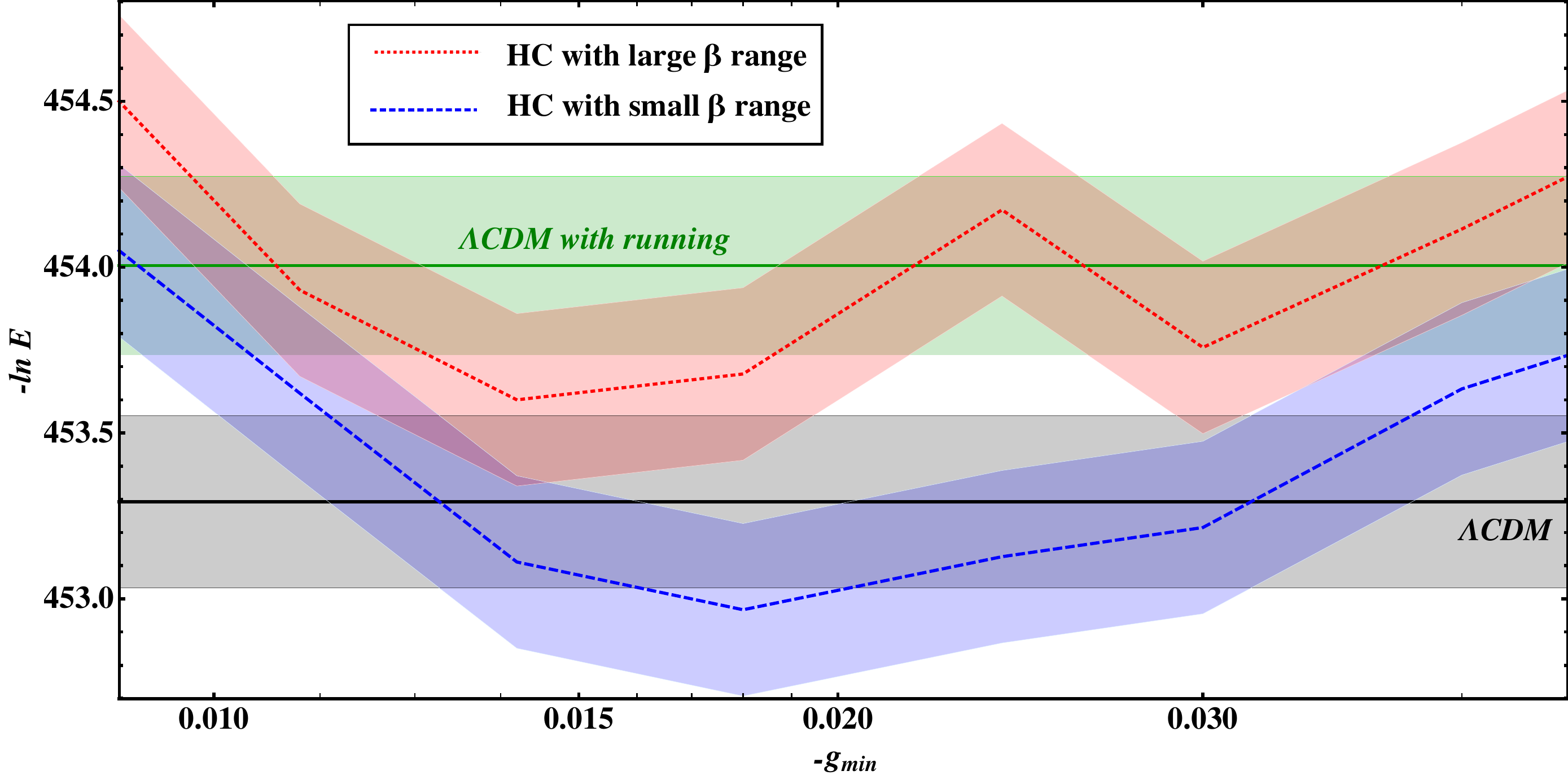}
\caption{Bayesian Evidence using $l \ge 30$\label{simfig30} data only, where the perturbative expansion (\ref{power_hc}) can be trusted. Error is indicated by the shaded region around the lines.}
\end{center}
\end{figure}

The results for the Bayesian evidence are presented in 
Fig. \ref{simfig30} for $l \geq 30$, where 2-loop predictions  (\ref{power_hc}) can be trusted.
As a guide \cite{Trotta:2008qt}, a difference $\ln E<1$  is insignificant and $2.5 < \ln E< 5$ is strongly significant.
We see that  the difference between evidence for $\Lambda$CDM and HC predictions is insignificant, with marginal preference for HC, depending on the choice of priors.

{\em Model selection.}--- We would like now to examine whether we can use the data to rule out or in some of the models described by (\ref{action}). There are phenomenological and theoretical constraints that we need to satisfy. The phenomenological constraints are: the bound on the tensor-to-scalar ratio, $r \leq 0.125$,  should be satisfied, and the model should reproduce the observed values for the amplitude $\Delta_0^2$ and  $\ln \beta$. The theoretical prediction for the $r$ is \cite{McFadden:2009fg, McFadden:2010na, Kawai:2014vxa}, 
\be \label{r}
r = 32 \frac{1 + \sum_{M=1}^{\mN_{\Phi}} (1-8 \xi_{M})^2}{1 + 2 \mN_\psi + \mN_{\Phi}},
\ee
and the theoretical predictions for $\Delta_0^2$ and $\ln \beta$ are given in (\ref{match}-\ref{final_log}). 
In deriving (\ref{power_hc}) we used a 't Hooft large $N$ expansion and perturbation theory in $\geff^2$. We thus need to check that any solution of the phenomenological constraints is consistent 
with these theoretical assumptions.

There are a few universal properties of the 2-loop correction, $g q_*/q \ln |q/\beta gq_*|$. This term vanishes at large $q$, reflecting the fact that the QFTs we consider are superrenormalizable.
Its absolute value gradually increases till it reaches the local maximum $1/ e \beta$ at $q= e \beta |g| q_*$. 
At lower values of $l$ the 2-loop term changes sign and grows very fast as we go to lower multipoles becoming equal to one (same size as the 1-loop contribution) below $l=10$. Therefore, we should not trust these models below $l \sim 10$. In fact, one should  even be cautious in using the 2-loop approximation for $l$'s lower than 35.
While the overall magnitude of the 2-loop term is small up until $l=10$ this happens due to large cancellation between the $f_1$ and the $f_2$ term in (\ref{2loop}). We will use as an  indicator of the reliability of perturbation theory the size of $f_1 \geff^2 \ln \geff^2$.



Let us consider gauge theory coupled to a large number, $\mN_\Phi$, of  non-mimimal scalars, all with the same non-minimality parameter $\xi$ and the same quartic coupling $\lambda$.
For sufficiently large $\mN_\Phi$, the scalar-to-tensor ratio (\ref{r}) becomes
\be \label{r2}
r = 32 (1- 8 \xi)^2
\ee
and the bound on $r$ implies,
 $|1- 8 \xi| \le 0.061$,
where the equality holds when $r=0.12.$ Choosing a value of $\xi$, then the observational values of $\Delta_0^2$ and $\ln \beta$ give two equations, which can always be solved to determine 
$N$ and $\mN_\Phi$. For example, if we choose $\xi=0.133$, which correspond to $r=0.12$,  and take $\lambda=1$ the solution to the two constraints is 
\be \label{sol}
N=2995, \qquad \mN_\Phi=23255.
\ee
This solution satisfies the theoretical constraints: firstly,  $N^2 \gg \mN_\Phi$, so 
the large $N$ expansion is justified and secondly,  the effective coupling remains small for all momenta seen by Planck, $3.3 \times 10^{-4} \le \geff^2(q)\le 0.41$. 
For this solution however $f_1 \approx  -8 (1-48 \xi^2)/(1-8 \xi)^2 \approx -11 $ and  $f_1 \geff^2 \ln \geff^2 \approx 1$ when $l=35$, so we should not trust the perturbative expansion below around $l \approx 35$.
 

{\em Conclusions.}---  We showed that holographic models based on three-dimensional perturbative QFT are capable of explaining the CMB data and are competitive to $\Lambda$CDM model.  However, at very low multipoles (roughly $l<30$), the perturbative expansion breaks down and in this regime the prediction of the theory cannot be trusted.  
The data are consistent with the dual theory being gauge theory coupled to a large number of nearly conformal scalars with a quartic interaction. It would be interesting to further analyze these models in order to extract other properties that may be testable against observations. In particular, non-perturbative methods (such as putting the dual QFT on a lattice) can be used to reliably model the very low multipoles, which may potentially explain the apparent large angle anomalies in the CMB sky (e.g., \cite{Ade:2013nlj}).

\begin{acknowledgments}
{\em Acknowledgments,}--- We would like to thank Raphael Flauger for collaboration at early stages of this work.
K.S. is supported in part by the Science and Technology Facilities Council (Consolidated Grant ``Exploring the Limits of the Standard Model and Beyond'').
K.S. would like to thank GGI in Florence for hospitality during the final stages of this work. 
N.A. and E.G. were supported in part by Perimeter Institute for Theoretical Physics. Research at Perimeter Institute is supported by the 
Government of Canada through the Department of Innovation, Science and Economic Development Canada and by the Province of Ontario through 
the Ministry of Research, Innovation and Science.
We acknowledge the use of the Legacy Archive for Microwave Background Data Analysis (LAMBDA), part of the High Energy Astrophysics Science Archive 
Center (HEASARC). HEASARC/LAMBDA is a service of the Astrophysics Science Division at the NASA Goddard Space Flight Center.
L.D.R. is partially supported by the ``Angelo Della Riccia'' foundation. The work of C.C. was supported in part by a {\it The Leverhulme Trust Visiting Professorship} 
at the STAG Research Centre and Mathematical Sciences, University of Southampton.
\end{acknowledgments}

\appendix

\section{Appendix: $\< TT \>$ at 2-loops}

The holographic formula for the power spectrum reads,
\be \label{power_sp_h}
\Delta^2_{{\cal R}}(q) =  -\frac{q^3}{4  \pi^2} 
\frac{1}{{\rm Im} \<\!\<T(q)T(-q)\>\!\>}  , 
\ee
where  $T=T^i_i$ is the trace of the energy momentum tensor
and  the double bracket notation indicates that the momentum conserving delta function (times $(2 \pi)^3$) has been removed. The imaginary part in (\ref{power_sp_h}) is taken after analytic continuation 
\be \label{analytic}
q \to -i q, N \to -i N, 
\ee
where $q$ is the magnitude of the momentum.  
This formula was derived in \cite{McFadden:2009fg} using the domain-wall/cosmology correspondence and also follows from the wave-function of the universe approach.
There is a similar holographic formula for the tensor power spectrum involving the transverse traceless part of the 2-point function of $T_{ij}$.

This class of theories we consider has the important property 
that if one promotes $g_{YM}^2$ to a new field that transforms appropriately under conformal transformation, the theory becomes conformally invariant \cite{Jevicki:1998ub, Kanitscheider:2008kd}. We say that the theory has a ``generalized conformal structure''. This is not a bona fide symmetry of the theory but, nevertheless, controls many of its properties. 
The generalised conformal structure implies that the 2-point function of the energy momentum tensor, to leading order in the large $N$ limit  (planar diagrams), is given by 
\bea
\<\!\< T(q) T(-q) \>\!\> = q^3 N^2 f(\geff^2),
\eea
where $\geff^2=\gYM^2 N/q$ is the effective dimensionless 't Hooft coupling constant and $f(\geff^2)$ is function of $\geff^2$ \cite{Kanitscheider:2008kd}. The overall factor of $q^3$ reflects the fact that the energy momentum has dimension 3 in three dimensions and the overall factor of $N^2$ is because this is the leading order term in the large $N$ limit.
Under the analytic continuation (\ref{analytic}),
\be
\geff^2(q) \to \geff^2(q), \quad N^2 q^3 \to -i N^2 q^3
\ee
and therefore for this class of theories, 
\be \label{power_sp3}
\Delta^2_{{\cal R}}(q) = \frac{1}{4 \pi^2 N^2 f(g_{eff}^2)},
\ee
which is the formula we used in the main text.

Perturbation theory is valid when $g_{\rm eff}^2 \ll 1$.  In the perturbative regime, the function  $f(\geff^2)$ is given by
\be \label{app:2loop}
f(\geff^2) = f_0 \left( 1 - f_1 \, \geff^2 \ln \geff^2 + f_2 \, \geff^2 + O(\geff^4)  \right).
\ee
The function $f_0$ is determined by a 1-loop computation, while $f_1$ and $f_2$ come from 2-loops. The presence of the logarithm is due to UV and IR divergences in the computation of the 2-point function of the energy momentum tensor. 

The 1-loop computation was done in \cite{McFadden:2009fg, McFadden:2010na} and we summarize the 2-loop computation of \cite{CDS} here. 
Since this a gauge theory we first need to gauge fix. The gauged fixed action is
\bea
S &=& \frac{1}{\gYM^2} \int d^3 x \, \tr \left[ \frac{1}{2} F_{ij} F^{ij} + (\partial^i A_i)^2 + 2 \partial^i \bar{c} \mD_i c \right. \nn \\
&& +  (\mD \Phi^M)^2  + 2 \bar{\psi}^L \mD \psi^L + 2 \sqrt{2} \, \mu_{M L_1 L_2} \Phi^M \bar{\psi}^{L_1} \psi^{L_2}  \nn \\
&& \left. + \frac{1}{6} \lambda_{M_1 M_2 M_3 M_4} \Phi^{M_1} \Phi^{M_2} \Phi^{M_3} \Phi^{M_4} \right] .
\eea
where $b, c$ is the ghost sector.
The energy-momentum tensor is obtained by coupling the theory to a background metric $g_{ij}$ and then using $T_{ij} = (2/\sqrt{g}) (\delta S/\delta g^{ij}) |_{g_{ij}=\delta_{ij}}$. 
This procedure defines a unique energy momentum tensor, except that there is a choice of how to couple the scalars to gravity. One may include the non-minimal coupling
$1/(2 \gYM^2) \sum_M \xi_M R (\Phi^M)^2$ in the action, where the sum is over all scalars and different scalars may have a different $\xi_M$. 
The variation of this term contributes an ``improvement term'' to the energy momentum tensor. 
If $\xi_M=0$ the scalar is a called a minimal scalar, while if $\xi_M=1/8$ it is a conformal scalar.  
The energy momentum obtained in this fashion is given by
\be
T_{ij} = T^A_{ij} + T^{g.f.}_{ij} + T^{gh}_{ij} + T^{\psi}_{ij} + T^{\Phi}_{ij} + T^{Y}_{ij},
\ee
where the different terms denote, respectively, the contribution of the gauge fields, the gauge-fixing term , the ghost sector, the fermions, the scalars and the Yukawa interactions. These are explicitly given by  
\bea \label{T}
T^{A}_{ij} &=& \frac{1}{\gYM^2} \tr \left[ F_{ik} F_{jk} - \delta_{ij} \frac{1}{4} F_{kl} F_{kl} \right] , \\
T^{g.f.}_{ij} &=& \frac{1}{ \gYM^2} \tr \left[ A_i \partial_j \left( \partial_k A_k\right)  + A_j \partial_i \left( \partial_k A_k\right) \right. \nonumber \\
&&\left. - \delta_{ij} \left( A_k\ \partial_k \partial_l A_l + \frac{1}{2} (\partial_k A_k) (\partial_l A_l) \right)\right] , \nn \\
T^{gh}_{ij} &=& \frac{1}{\gYM^2} \tr \left[ \partial_i \bar{c} \mD_j c + \partial_j \bar{c} \mD_i c - \delta_{ij} \partial_k \bar{c} \mD_k c\right] , \nn \\
T^{\psi}_{ij} &=& \frac{1}{\gYM^2} \tr \left[ \frac{1}{2} \bar{\psi}^L \gamma_{(i} \oa{\mD}_{j)} \psi^L - \delta_{ij} \frac{1}{2} \bar{\psi}^L \gamma_k \oa{\mD}_k \psi^L \right] ,\nn \\
T^{\Phi}_{ij} &=& \frac{1}{\gYM^2} \tr \left[ \mD_i \Phi^M \mD_j \Phi^M - \delta_{ij} \left( \frac{1}{2} (\mD \Phi^M)^2 \right. \right. \nonumber \\
&& \left. \qquad \qquad  + \frac{1}{4!}  \lambda_{M_1 M_2 M_3 M_4} \Phi^{M_1} \Phi^{M_2} \Phi^{M_3} \Phi^{M_4}  \right)  \nonumber \\
&& \left. \qquad \qquad + \, \xi_M \left( \delta_{ij} \partial^2 - \partial_i \partial_j\right) (\Phi^M)^2\right] , \nn \\
T^{Y}_{ij} &=& \frac{1}{\gYM^2} \tr \left[ - \delta_{ij} \, \, \mu_{M L_1 L_2} \Phi^M \bar{\psi}^{L_1} \psi^{L_2} \right] . \nn
\eea

\begin{figure}[h]
\centering
\includegraphics[scale=1]{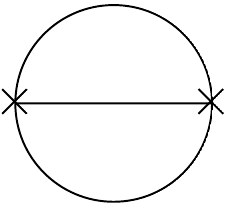} \hspace{0.5cm}
\includegraphics[scale=1]{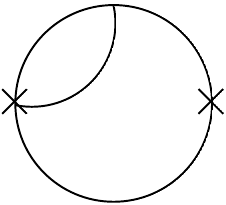} \hspace{0.5cm}
\includegraphics[scale=1]{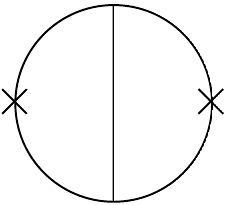} \hspace{0.5cm}
\includegraphics[scale=1]{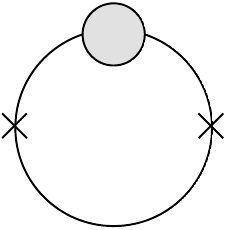} \hspace{0.5cm}
\includegraphics[scale=1]{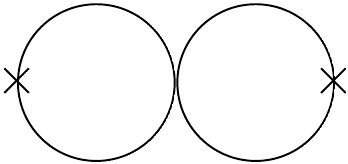}
\caption{2-loop diagrams contributing to $\< TT \>$. The  blob in the fourth diagram represents an insertion of a 1-loop self-energy. \label{Fig.TT2L}}
\end{figure}

The topology of the 2-loop diagrams that need to be computed is given in Fig. \ref{Fig.TT2L}.
The 2-point function has a UV divergence that can be cancelled by the counterterm
\be
a_{CT} \int d^3 x \sqrt{g} R,
\ee
where $R$ is the curvature scalar of the background metric $g_{ij}$, with appropriately chosen $a_{CT}$. As usual, this process introduces 
a renormalization scale $\mu$, which leads to scheme dependence. The correlator also has an IR divergence (unless all scalars are minimal), 
which we regulated with an IR cut-off, $\mu_*$. This leads to the following result for $\ln \beta$,
\be
\ln \beta=\ln \frac{1}{|g|} -\frac{a_0}{f_1} - \left(\ln \frac{q_*}{\mu} + \frac{64/\pi^2}{f_1 \mN_{(B)} }\Sigma_\Phi \ln \frac{\mu}{\mu_*}\right),
\ee
where we have made use of the definition of $g$, $g q_* = f_1 \gYM^2 N$. The renormalization scale $\mu$ is arbitrary and so is the pivot scale $q_*$.
We fix this scheme dependence by setting, $\mu=q_*$. An alternative scheme is to set $\mu$ to the (inverse of the) smallest scale in the data (i.e. equal to  0.17 Mpc${}^{-1}$).
We have checked that the results we present in the main text are not sensitive to the choice of scheme, except possibly at very low $l$'s. As argued in the main text, this is precisely the regime 
where one should not trust the perturbative computation. Regarding the IR divergence now. It was argued in \cite{Jackiw:1980kv, Appelquist:1981vg} that 
superrenormalizable theories with a dimensionful coupling constant (as in our case)
are non-perturbative IR finite, with $\gYM^2$ providing the IR cut-off . We therefore set $\mu_*= c \gYM^2$, where $c$ is a number that can only be determined non-perturbatively.  This leads to our final formula for $\ln \beta$,
\be \label{app:final_log}
\ln \beta=\ln \frac{1}{|g|} -\frac{a_0}{f_1} - \frac{64/\pi^2}{ f_1 \mN_{(B)} }\Sigma_\Phi \ln \frac{N f_1}{c g}
\ee
In the main text we set $c=1$ but we also checked that the results do not change qualitatively if we change $c$. An alternative way to deal with the IR issues is to get $\mu_*$ equal to the (inverse of the) largest scale in the data (i.e. $1.4 \times 10^{-4}$ Mpc${}^{-1}$). As in the case of scheme dependence, we have checked that the results are not very sensitive to how we treat $\mu_*$, except possibly at very low $l$s.

%


\end{document}